\documentclass{article}
\usepackage{spconf,amsmath,graphicx}
\usepackage{amsfonts}
\usepackage{amssymb}
\usepackage{bm}
\usepackage{color, soul}
\usepackage{stfloats}
\usepackage[numbers,sort&compress]{natbib}
\usepackage[amsmath,thmmarks]{ntheorem}
\usepackage{theorem}
\usepackage{subfigure}
\usepackage{ntheorem}
\usepackage{url}
\usepackage{hyperref}
\usepackage{float}
\usepackage{mathrsfs}
\usepackage{algorithm}
\usepackage{algorithmic}
\usepackage{subfigure}
\usepackage{mathrsfs}
\usepackage{enumerate}
\usepackage{enumitem}
\setlist[enumerate,1]{label=(\arabic*).,font=\textup,
leftmargin=7mm,labelsep=1.5mm,topsep=0mm,itemsep=-0.8mm}
\setlist[enumerate,2]{label=(\alph*).,font=\textup,
 leftmargin=7mm,labelsep=1.5mm,topsep=-0.8mm,itemsep=-0.8mm}

\title{Gridless Line Spectral Estimation with Multiple Measurement Vector via Variational Bayesian Inference}
\name{Qi Zhang$^+$, Jiang Zhu$^+$, Peter Gerstoft$^*$, Mihai-Alin Badiu$^{\dagger \diamond}$ and Zhiwei Xu$^+$}
\address{$^+$Ocean College, Zhejiang University,  Zhoushan, CHINA\\$^*$Electrical and Computer Engineering, University of California, San Diego, USA\\$^{\dagger}$Department of Engineering Science, University of Oxford, UK\\$^{\diamond}$Department of Electronic Systems, Aalborg University, Denmark}
\begin{document}
\maketitle
\begin{abstract}
Line spectral estimation (LSE) from multi snapshot samples is studied utilizing the variational Bayesian methods. Motivated by the recently proposed variational line spectral estimation (VALSE) method for a single snapshot, we develop the multisnapshot VALSE (MVALSE) for multi snapshot scenarios, which is important for array processing. The MVALSE shares the advantages of the VALSE method, such as automatically estimating the model order, noise variance and weight variance, closed-form updates of the posterior probability density function (PDF) of the frequencies. By using multiple snapshots, MVALSE improves the recovery performance and it encodes the prior distribution naturally. Finally, numerical results demonstrate the competitive performance of the MVALSE compared to state-of-the-art methods.
\end{abstract}
%\begin{keywords}
\section{Introduction}
\label{sec:intro}
Line spectral estimation (LSE) is a fundamental problem in signal processing \cite{Stoica} with many applications including channel estimation \cite{Hansen0}, direction of arrival estimation \cite{Huanglei1,Huanglei2}. Classical methods such as MUSIC and ESPRIT \cite{Schmidt, Roy} utilize the covariance matrix to estimate the frequencies. The estimation accuracy of these methods can degrade significantly if the model order is unknown.

Recently, compressed sensing (CS) based methods have drawn a great deal of attention. By constructing a dictionary matrix with the frequencies being restricted onto the grid, LSE is formulated as a sparse signal recovery problem, and several algorithms such as $\ell_1$ based optimization and sparse iterative covariance-based estimation (SPICE) are proposed \cite{Malioutov, Stoica2, Stoica3, Stoica4}. However, the above on grid sparse methods suffer model mismatch when the true frequencies do not lie on the grid \cite{Chi1}. To resolve the grid mismatch problem, off-grid based methods are proposed \cite{Tang}.
The off-grid based methods first perform the sparse estimation on the grid, and then refine the estimation without restricting on the grid \cite{Fang, Mamandipoor}. Meanwhile, gridless sparse methods which directly operate in the continuous domain are also proposed, such as gridless SPICE (GLS), sparse and parametric approach (SPA) for single measurement vector (SMV) case and atom based method such as SPA and other methods for the multiple measurement vector (MMV) case \cite{Chi, Yang1, Yang2, Yang3, Hansen}.

This paper develops the multisnapshot VALSE (MVALSE) algorithm by extending the gridless variational line spectral estimation (VALSE) algorithm in the SMV case \cite{Badiu} to the MMV case, which is especially important for array signal processing \cite{Gerstoft}. In addition, the prior information may be available, for example, in frequency estimation where the received frequency differ with the known transmitted frequency due to Doppler, sequential estimation where the past results is acted as the prior information, etc. Here a von Mises prior distribution is encoded into the MVALSE algorithm, and the performance of MVLASE improves with the correct prior distribution.
\section{Problem Setup}
The measurement $\mathbf Y\in\mathbb{C}^{M\times L}$ for $M$ sensors and $L$ observations is given by
\begin{small}
\begin{align}\label{signal-generate}
{\mathbf Y} = \sum\limits_{k=1}^{K}{\mathbf a}({\widetilde\theta}_k){\widetilde{\mathbf w}}_k^{\rm T}+{\widetilde{\mathbf U}},
\end{align}
\end{small}where $K$ is the number of spectral components which is generally unknown, ${\widetilde{\mathbf w}}_k \in {\mathbb C}^{L\times 1}$ and $\widetilde\theta_k\in[-\pi,\pi)$ are the complex weight and frequency of the $k$th component, respectively, ${\mathbf a}(\widetilde\theta_k)=\left[1,{\rm e}^{{\rm j}{\widetilde\theta}_k},\cdots,{\rm e}^{{\rm j}(M-1){\widetilde\theta}_k}\right]^{\rm T}$, ${\widetilde{\mathbf U}}$ is the additive white Gaussian noise (AWGN).

Since the number of spectral components $K$ is unknown, the measurement $\mathbf Y$ is assumed to consist of a superposition of $N$ components with $N>K$ \cite{Badiu} , i.e.,
\begin{small}
\begin{align}\label{psuedomodel}
{\mathbf Y} = \sum\limits_{i=1}^{N}{\mathbf a}(\theta_i){\mathbf w}_i^{\rm T}+{\mathbf U} = {\mathbf A}{\mathbf W} + {\mathbf U},
\end{align}
\end{small}where ${\mathbf A} = [{\mathbf a}(\theta_1),\cdots,{\mathbf a}(\theta_N)]\in\mathbb{C}^{M\times{N}}$, the $i$th column of $\mathbf A$ is ${\mathbf a}(\theta_i)=[1,{\rm e}^{{\rm j}\theta_i},\cdots,{\rm e}^{{\rm j}(M-1)\theta_i}]^{\rm T}$, ${\mathbf w}_i^{\rm T}$ denote the $i$th row of $\mathbf W \in\mathbb{C}^{N\times{L}}$, the elements of the noise ${\mathbf U} \in {\mathbb C}^{M\times L}$ are i.i.d. and $U_{ij}\sim {\mathcal {CN}}(U_{ij};0,\nu)$, where ${\mathcal {CN}}({\mathbf x};{\boldsymbol \mu},{\boldsymbol \Sigma})$ is the complex normal distribution of ${\mathbf x}$ with mean ${\boldsymbol \mu}$ and covariance ${\boldsymbol \Sigma}$. In addition, binary hidden variables ${\mathbf s}\in {\mathbb R}^N$ are introduced, and the probability distribution is
$p({\mathbf s};\rho)=\prod_{i=1}^Np(s_i;\rho)$, where $s_i\in\{0,1\}$ and
\begin{small}
\begin{align}\label{pmfs}
p(s_i;\rho) = \rho^{s_i}(1-\rho)^{(1-s_i)}.
\end{align}
\end{small}We assume $p({\mathbf W}|{\mathbf s};\tau)=\prod_{i=1}^Np({\mathbf w}_i|s_i;\tau)$, where $p({\mathbf w}_i|s_i;\tau)$ follows a Bernoulli-Gaussian distribution
\begin{small}
\begin{align}\label{pdfws}
p({\mathbf w_i}|s_i;\tau) = (1 - s_i){\delta}({\mathbf w_i}) + s_i{\mathcal {CN}}({\mathbf w_i};{\mathbf 0},\tau{\mathbf I}),
\end{align}
\end{small}where ${\delta}(\cdot)$ is the Dirac delta function. From (\ref{pmfs}) and (\ref{pdfws}), it can be seen that $\rho$ controls the probability of the $i$th component being active. The prior distribution $p({\boldsymbol \theta})$ of the frequency ${\boldsymbol \theta} = [\theta_1,...,\theta_N]^{\rm T}$ is $p({\boldsymbol \theta}) = \begin{matrix} \prod_{i=1}^N p(\theta_i) \end{matrix}$, where $p(\theta_i)$ is encoded through the von Mises distribution \cite[p.~36]{Direc}
\begin{small}
\begin{align}\label{prior_theta}
p(\theta_i) = {\mathcal {VM}}(\theta_i;\mu_{0,i},\kappa_{0,i})= \frac{1}{2\pi{I_0}(\kappa_{0,i})}{\rm e}^{\kappa_{0,i}{\cos(\theta-\mu_{0,i})}},
\end{align}
\end{small}where $\mu_{0,i}$ is the mean direction,
$\kappa_{0,i}$ is the measure of concentration, $I_p(\cdot)$ is the modified Bessel function of the first kind and the order $p$ \cite[p.~348]{Direc}.
Note that $\kappa_{0,i}=0$ corresponds to the uninformative prior distribution $p(\theta_i) = {1}/({2\pi})$ \cite{Badiu}.

For measurement model
(\ref{psuedomodel}), the likelihood $p({\mathbf Y}|{\mathbf A}{\mathbf W};\nu)$ is
\begin{small}
\begin{align}
p({\mathbf Y}|{\mathbf A}{\mathbf W};\nu)=\prod\limits_{i,j}{\mathcal {CN}}(Y_{ij};[{\mathbf A}{\mathbf W}]_{ij},\nu),
\end{align}
\end{small}where $[\cdot]_{ij}$ is the $(i,j)$th element.
Let ${\boldsymbol \beta} = \{\nu,~\rho,~\tau\}$ and ${\boldsymbol \Phi}=\{{\boldsymbol \theta}, {\mathbf W}, {\mathbf s}\}$ be the model and estimated parameters. Given the above statistical model, computing the maximum likelihood (ML) estimate of ${\boldsymbol \beta}$ and the maximum a posterior (MAP) estimate of ${\boldsymbol \Phi}$ are intractable. Thus an iterative algorithm is designed.

\section{The MVALSE Algorithm}
Here a mean field variational Bayes method is proposed. Full details are in \cite{Qijournal}, and here we summarize the main results. For any assumed PDF $q({\boldsymbol \Phi}|{\mathbf Y})$, the marginal likelihood $p({\mathbf Y};{\boldsymbol\beta})$ is \cite{Murphy}
\begin{small}
\begin{align}\label{DL}
\ln p({\mathbf Y};{\boldsymbol\beta})= {\rm{KL}}\left(q({\boldsymbol \Phi}|{\mathbf Y})||p({\boldsymbol \Phi}|{\mathbf Y})\right) + {\mathcal L}(q({\boldsymbol \Phi}|{\mathbf Y});{\boldsymbol\beta}),
\end{align}
\end{small}where ${\rm{KL}}(\cdot||\cdot)$ is the Kullback-Leibler divergence. Since ${\rm{KL}}(\cdot||\cdot)\geq 0$, it can be seen that ${\mathcal L}(q({\boldsymbol \Phi}|{\mathbf Y});{\boldsymbol\beta})$ provides a lower bound on the marginal likelihood. Our goal is to compute
posterior distribution approximations $q({\boldsymbol \Phi}|{\mathbf Y})$ by maximizing the lower bound ${\mathcal L}(q({\boldsymbol \Phi}|{\mathbf Y}))$ with $q({\boldsymbol \Phi}|{\mathbf Y})$ factored as
\begin{small}
\begin{align}
q({\boldsymbol \Phi}|{\mathbf Y}) &= \prod_{i=1}^Nq(\theta_i|{\mathbf Y})q({\mathbf W},{\mathbf s}|{\mathbf Y})\\
&=\prod_{i=1}^Nq(\theta_i|{\mathbf Y})q({\mathbf W}|{\mathbf Y},{\mathbf s})\delta({\mathbf s}-{\widehat{\mathbf s}}).\label{postpdf}
\end{align}
\end{small}Given (\ref{postpdf}), the frequencies ${\boldsymbol \theta}$ are estimated as
\begin{small}
\begin{subequations}\label{ahat}
\begin{align}
&\widehat{\theta}_i = {\rm arg}({\rm E}_{q({\theta_i|\mathbf Y})}[{\rm e}^{{\rm j}\theta_i}]),\label{ahata}\\
&\widehat{\mathbf a}_i = {\rm E}_{q({\theta_i|{\mathbf Y}})}[{\mathbf a}({\theta}_i)],~i\in\{1,...,N\},\label{ahatb}
\end{align}
\end{subequations}
\end{small}where ${\rm arg}(\cdot)$ returns the angle.
Besides, the mean and covariance estimates of the weights are calculated as
\begin{small}
\begin{subequations}\label{w_est}
\begin{align}
&\widehat{\mathbf w}_i = {\rm E}_{q({\mathbf W}|{\mathbf Y})}[{\mathbf w}_i],\\
&\widehat{\mathbf C}_{i,j} = {\rm E}_{q({\mathbf W}|{\mathbf Y})}[{\mathbf w}_i{\mathbf w}_j^{\rm H}] - \widehat{\mathbf w}_i\widehat{\mathbf w}_j^{\rm H},~i,j\in\{1,...,N\},
\end{align}
\end{subequations}
\end{small}where $q(\mathbf {W|Y}) \triangleq q({\mathbf W}|{\mathbf Y},\widehat{\mathbf s})$.
The set of indices of the non-zero components of ${\mathbf s}$ and the estimated model order are
\begin{small}
\begin{align}\notag
{\mathcal S} = \{i|1\leq i\leq N,s_i = 1\}\indent{\rm and }\indent\widehat{K} = |\widehat{\mathcal S}|.
\end{align}
\end{small}
The reconstructed line spectral signal $\widehat{\mathbf X}$ is
\begin{small}
\begin{align}\notag
\widehat{\mathbf X} = \sum_{i\in{\widehat{\mathcal S}}}\widehat{\mathbf a}_i\widehat{\mathbf w}_i^{\rm T}.
\end{align}
\end{small}Let ${\mathbf z}=(\theta_1,\cdots,\theta_N,({\mathbf W},{\mathbf s}))$ be the set of all latent variables. The posterior approximation $q({\mathbf z}_d)$ of each latent variable ${\mathbf z}_d,~d=1,\cdots,N+1$ is found using \cite[pp. 735, eq. (21.25)]{Murphy}
\begin{small}
\begin{align}\label{upexpression}
\ln q({\mathbf z}_d)={\rm E}_{q({{\mathbf z}\setminus{\mathbf z}_d})}[\ln p({\mathbf Y},{\mathbf z})]+{\rm const},
\end{align}
\end{small}where the expectation is with respect to all the variables ${\mathbf z}$ except ${\mathbf z}_d$.

Maximizing ${\mathcal L}(q({\mathbf z}|{\mathbf Y}))$ with respect to all the factors is also intractable. Similar to the Gauss-Seidel method \cite{Bertsekas},
we optimize $\mathcal L$ over each factor $q({\mathbf z}_i|{\mathbf Y})$, $i=1,\cdots,N+1$ separately with the others being fixed iteratively. In the following, we detail the procedure.

\subsection{Inferring the frequencies}\label{yita}
According to (\ref{upexpression}), the posterior approximation $q(\theta_i|{\mathbf Y})$ of the $i$th frequency is \cite{Qijournal}
\begin{small}
\begin{align}\label{pdf-q}
q(\theta_i|\mathbf Y)\propto \underbrace{p(\theta_i)}_{(a)}\underbrace{\exp({\rm Re}(\boldsymbol \eta_i^{\rm H}\mathbf a(\theta_i)))}_{(b)},
\end{align}
\end{small}where $\propto$ is identity up to a normalizing constant and the complex vector $\boldsymbol\eta_i$ is given by
\begin{small}
\begin{align}
\boldsymbol\eta_i = {2}/{\nu}({\mathbf Y{\widehat{\mathbf w}_i}^*} - \sum_{l\in\widehat{\mathcal S} \backslash \{i\}}\widehat{\mathbf a}_l({\rm tr}({\widehat{\mathbf C}}_{l,i}) + {\widehat{\mathbf w}}_i^{\rm H}{\widehat{\mathbf w}}_l))
\end{align}
\end{small}when $i\in\widehat{\mathcal S}$ and $(\cdot)^*$ is the conjugate operation, and $\boldsymbol \eta_i = \boldsymbol 0$ otherwise.

For each iteration, (\ref{ahatb}) needs to be computed to obtain the approximate posterior distribution of $\mathbf W$, as shown in the next subsection. Therefore, we approximate (\ref{pdf-q}) as a von Mises distribution, which gives a closed-form approximation of ${\rm E}_{q({\theta_i}|{\mathbf Y})}[{\mathbf a}(\theta_i)]$, see \cite{Badiu, Direc} for further details about von Mises PDFs.

In (\ref{pdf-q}), the correspondence between the prior $(a)$ and the likelihood $(b)$ is unknown. Therefore, part $(b)$ is first approximated as a von Mises distribution ${\mathcal {VM}}(\theta_i;\mu_{i},\kappa_i)$, then the prior is chosen as $p(\theta_j)$, where $j$ is $\underset{j}{\operatorname{argmax}}|\kappa_i{\rm e}^{{\rm j}\mu_{i}}+\kappa_{0,j}{\rm e}^{{\rm j}\mu_{0,j}}|$. After $q(\theta_i|\mathbf Y)$ is approximated as a von Mises distribution, a Newton step is added to refine the mean direction and the concentration parameter of the approximated von Mises distribution. For the second frequency, the prior can be similarly chosen from the set $\{p(\theta_i)\}_{i=1}^N$ with the first selected prior being removed. For the other frequencies, the steps follow similarly.
\subsection{Inferring the weights and support}\label{W-and-C}
From (\ref{postpdf}), the posterior approximation $q({\mathbf W},{\mathbf s}|{\mathbf Y})$  can be factored as the product of $q({\mathbf W}|{\mathbf Y},{\mathbf s})$ and $\delta({\mathbf s}-{\widehat{\mathbf s}})$. Define the matrix $\mathbf J$ with elements
\begin{small}
\begin{align}
J_{ii} = M,\quad J_{ij} = {\widehat{\mathbf a}}^{\rm H}_i{\widehat{\mathbf a}}_j,~i,~j = 1,...,N,~j\neq i,
\end{align}
\end{small}and ${\mathbf H}=\widehat{\mathbf A}^{\rm H}{\mathbf Y}$.
Let $\mathbf H_{\mathcal S}$ be the submatrix by choosing the rows of $\mathbf H$ indexed by ${\mathcal S}$ and $\mathbf J_{\mathcal S}$ be the submatrix by choosing both the rows and columns of $\mathbf J$ indexed by $\mathcal S$. According to (\ref{upexpression}), $q({\mathbf W},{\mathbf s}|{\mathbf Y})$ is updated as \cite{Qijournal}
\begin{small}
\begin{align}
&q({\mathbf W}|{\mathbf Y}) = {\mathcal {CN}}({\rm vec}({\mathbf W}_{\widehat{\mathcal S}});{\rm vec}({\widehat{\mathbf W}}_{\widehat{\mathcal S}}),{\widehat{\mathbf C}}_{\widehat{\mathcal S}})\prod_{i\not\in\widehat{\mathcal S}}\delta(\mathbf w_i),\\
&\widehat{\mathbf s} = \underset{\mathbf s}{\operatorname{argmax}}~\ln Z({\mathbf s})\label{findings},
\end{align}
\end{small}where
\begin{small}
\begin{subequations}\label{W-C-1}
\begin{align}
&\widehat{\mathbf W}_{\widehat{\mathcal S}} = \nu^{-1}\widehat{\mathbf C}_{\widehat{\mathcal S},0}\mathbf H_{\widehat{\mathcal S}},\label{What}\\
&\widehat{\mathbf C}_{\widehat{\mathcal S}}=\nu({\mathbf J}_{\widehat{\mathcal S}}+\frac{\nu}{\tau}{\mathbf I}_{|\widehat{\mathcal S}|})^{-1}\otimes{\mathbf I_L}\triangleq {\widehat{\mathbf C}_{\widehat{\mathcal S},0}}\otimes{\mathbf I_L},\label{defCshat0}\\
&\ln Z({\mathbf s}) = -L\ln\det(\mathbf J_{\mathcal S}+\frac{\nu}{\tau}\mathbf I_{|\mathcal S|})+||\mathbf s||_0\ln\frac{\rho}{1-\rho}\notag\\
&+\nu^{-1}{\rm tr}(\mathbf H_{\mathcal S}^{\rm H}(\mathbf J_{\mathcal S}+\frac{\nu}{\tau}{\mathbf I}_{|\mathcal S|})^{-1}{\mathbf H}_{\mathcal S})+||\mathbf s||_0L\ln\frac{\nu}{\tau}+{\rm const}.\label{lnZ}
\end{align}
\end{subequations}
\end{small}

The computation cost of enumerative solving (\ref{findings}) to find the globally optimal binary sequence $\mathbf s$ is $O(2^N)$, which is impractical for typical values of $N$. Here a greedy iterative search strategy is adopted \cite{Badiu}. For a given $\widehat{\mathbf s}$, we update it as follows: For each $k=1,\cdots,N$, calculate $\Delta_k=\ln Z(\widehat{\mathbf s}^k)-\ln Z(\widehat{\mathbf s})$, where $\widehat{\mathbf s}^k$ is the same as $\widehat{\mathbf s}$ except that the $k$th element of $\widehat{\mathbf s}$ is flipped. Let $k^*=\underset{k}{\operatorname {argmax}}~\Delta_k$. If $\Delta_{k^*}>0$, we update $\widehat{\mathbf s}$ with the $k^*$th element flipped, and $\widehat{\mathbf s}$ is updated, otherwise $\widehat{\mathbf s}$ is kept, and the algorithm is terminated. In fact, $\Delta_k$ can be easily calculated and the details are referred to \cite{Qijournal}.
\subsection{Estimating the model parameters}
After updating the approximate distribution of the frequencies, weights and $\mathbf s$ as ${q}({\boldsymbol \Phi})$, the model parameters are updated as \cite{Qijournal}
\begin{small}
\begin{align}
\widehat{\nu} =& {||\mathbf Y-{\widehat{\mathbf A}}_{\widehat{\mathcal S}}{\widehat{\mathbf W}}^{\rm T}_{\widehat{\mathcal S}}||^2_{\rm F}}/({ML})+{{\rm tr}(\mathbf J_{\widehat{\mathcal S}}\widehat{{\mathbf C}}_{\widehat{\mathcal S},0})}/{M}\notag\\
&+\sum_{i\in\widehat{\mathcal S}}\sum_{g=1}^L|\widehat{w}_{gi}|^2({1}/{L}-{||\widehat{\mathbf a}_i||_2^2}/({LM})),\label{rou-hat}
\end{align}
\end{small}
\begin{small}
\begin{align}\label{beta-tau-hat}
\widehat{\rho} = \frac{||\widehat{\mathbf s}||_0}{N}~~{\rm and}~~\widehat{\tau} = \frac{{\rm tr}(\widehat{\mathbf W}_{\widehat{\mathcal S}}\widehat{\mathbf W}^{\rm H}_{\widehat{\mathcal S}})+L{\rm tr}(\widehat{{\mathbf C}}_{\widehat{\mathcal S},0})}{L||\widehat{\mathbf s}||_0}.
\end{align}
\end{small}
\subsection{Initialization and Algorithm}
Initialization is important for the MVALSE algorithm and the details for initializing both the approximate distributions and model parameters are presented in \cite{Qijournal}.
To sum up, the MVALSE algorithm is summarized as Algorithm \ref{VALSEMMV}.
\begin{algorithm}[ht]
\caption{Outline of MVALSE algorithm.}\label{VALSEMMV}
\textbf{Input:}~~Signal matrix $\mathbf Y$\\
\textbf{Output:}~~Model order $\widehat{K}$, frequencies $\widehat{\boldsymbol\theta}_{\widehat{\mathcal S}}$, complex weights $\widehat{\mathbf W}_{\widehat{\mathcal S}}$ and reconstructed signal $\widehat{\mathbf X}$
\begin{algorithmic}[1]
\STATE Initialize $\widehat{\nu},\widehat{\rho},\widehat{\tau}$~and~$q(\theta_i|{\mathbf Y}),i\in\{1,\cdots,N\}$; compute $\widehat{\mathbf a}_i$
\STATE \textbf{repeat}
\STATE ~~~~Update~$\widehat{\mathbf s},\widehat{\mathbf W}_{\widehat{\mathcal S}}~{\rm and}~\widehat{\mathbf C}_{\widehat{\mathcal S}}$ ({\rm Sec}.\ref{W-and-C})
\STATE ~~~~Update~$\widehat{\nu}$ (\ref{rou-hat}), $\widehat{\rho}$, $\widehat{\tau}$ (\ref{beta-tau-hat})
\STATE ~~~~Update~$\boldsymbol\eta_i$~and~$\widehat{\mathbf a}_i$ for all $i\in \widehat{\mathcal S}$ ({\rm Sec}.\ref{yita})
\STATE \textbf{until} stopping criterion
\STATE \textbf{return} $\widehat{K}$, $\widehat{\boldsymbol\theta}_{\widehat{\mathcal S}}$, $\widehat{\mathbf W}_{\widehat{\mathcal S}}$ and $\widehat{\mathbf X}$
\end{algorithmic}
\end{algorithm}
\subsection{Computation complexity analysis}
The complexity per iteration is dominated by the two steps \cite{Badiu}: the maximization of $\ln Z({\mathbf s})$ and the approximations of the posterior PDF
$q(\theta|{\mathbf Y})$ by mixtures of von Mises pdfs. For the MVALSE algorithm, the complexity of the two steps are $O(N^4 + N^3L)$ and $O(MN(M+N+L))$ \cite{Qijournal}. In conclusion, the dominant computational complexity of the MVALSE is $O[(N^4+N^3L)\times T]$ with $T$ being the number of iterations as $M$ is close to $N$.
\section{Simulation}
In this section, numerical experiments are conducted to evaluate the performance of the proposed algorithm. The frequencies are generated as follows: First, $K$ distributions are uniformly picked from $N$ von Mises distributions (\ref{prior_theta}) with $\mu_{0,i}=(2i-1-N)/(N+1)\pi$ and $\kappa_{0,i}=10^4$, $~i=1,\cdots,N$ without replacement. The frequencies $\{\theta_i\}_{i=1}^K$ are generated from the selected von Mises distribution and the minimum wrap-around distance is greater than $\triangle\theta=\frac{2\pi}{N}$. The elements of $\mathbf W$ are drawn i.i.d. from ${\mathcal {CN}}(0,1)$. Other parameters are: $K=3$, $N=20$, $M=20$, ${\rm SNR}=4$ dB, where signal-to-noise ratio (SNR) is ${\rm SNR} \triangleq 10{\rm log}(||\mathbf A(\widetilde{\boldsymbol\theta})\widetilde{\mathbf W}^{\rm T}||_{\rm F}^2/||\widetilde{\mathbf U}||_{\rm F}^2)$ \footnote{For the numerical experiment where $\kappa_{0,i}=10^4$ and $N=20$, straightforward calculation shows that the standard deviation of the von Mises distribution $\approx0.01$ and the distance between the adjacent frequencies is $\mu_{0,i+1}-\mu_{0,i}=0.3$. Thus the MVALSE with prior is almost a grid based method.}. The normalized mean square error (NMSE) of $\widehat{\mathbf Y}$ and $\widehat{\boldsymbol\theta}$ defined as ${\rm NMSE}(\widehat{\mathbf X}) \triangleq 10{\rm log}(||\widehat{\mathbf X}-{\mathbf A}(\widetilde{\boldsymbol\theta})\widetilde{\mathbf W}^{\rm T}||_{\rm F}^2/||{\mathbf A}(\widetilde{\boldsymbol\theta})\widetilde{\mathbf W}^{\rm T}||_{\rm F}^2)$ and ${\rm NMSE}(\widehat{\boldsymbol\theta}) \triangleq 10{\rm log}(||\widehat{\boldsymbol\theta} - \widetilde{\boldsymbol\theta}||_2^2/||\widetilde{\boldsymbol\theta}||_2^2)$, the correct model order estimated probability $P(\widehat{K}=K)$ are adopted as the performance metrics. In the case when the model order is overestimated such that $\widehat{K}>K$, the top $K$ elements of $\widehat{\boldsymbol{\kappa}}$ is chosen to calculate the NMSE of the frequency, where $\widehat{\kappa}_i$ is the concentration parameter of the posterior $q(\theta_i|{\mathbf Y})$. When $\widehat{K}<K$, the frequencies are filled with zeros to calculated the NMSE of the frequency. The Algorithm \ref{VALSEMMV} stops when  $||\widehat{\mathbf X}^{(t-1)} - \widehat{\mathbf X}^{(t)}||_2/||\widehat{\mathbf X}^{(t-1)}||_2 < 10^{-5}$ or $t > 200$, where $t$ is the number of iteration.

In addition, the SPA method \cite{Yang1}, the Newtonized orthogonal matching pursuit (NOMP) method \cite{Mamandipoor, Lin} and the Cram\'er-Rao bound (CRB) derived in \cite{Lin} are chosen for performance comparison. For the SPA approach, the denoised covariance matrix is obtained firstly and the MUSIC method is used to avoid frequency splitting phenomenon, where the MUSIC method is provided by MATLAB \emph{rootmusic} and the optimal sliding window $W$ is empirically found. We set $W=12$. For the NOMP method, the termination condition is set such that the probability of model order overestimate is $1\%$ \cite{Lin}. All results are averaged over $10^3$ Monte Carlo (MC) trials.
\begin{table}[h!t]
    \begin{center}
\caption{The empirical probability of $\widehat{K}>K$  of the algorithms.}\label{table}
    \begin{tabular}{|c|c|c|c|c|c|c|}
            \hline
            snapshots $L$&1&3&5&7\\ \hline
            MVALSE, prior & 33\% & 31\% & 1\%&0 \\ \hline
            MVALSE, noninfo. prior & 30\% & 23\% & 1\% & 0  \\ \hline
            NOMP & \multicolumn{4}{|c|}{1\%} \\ \hline
             \end{tabular}
    \end{center}
\end{table}

\begin{figure}\label{L}
  \center
  \subfigure[]{
    \label{L-X} %% label for first subfigure
    \includegraphics[width=60mm]{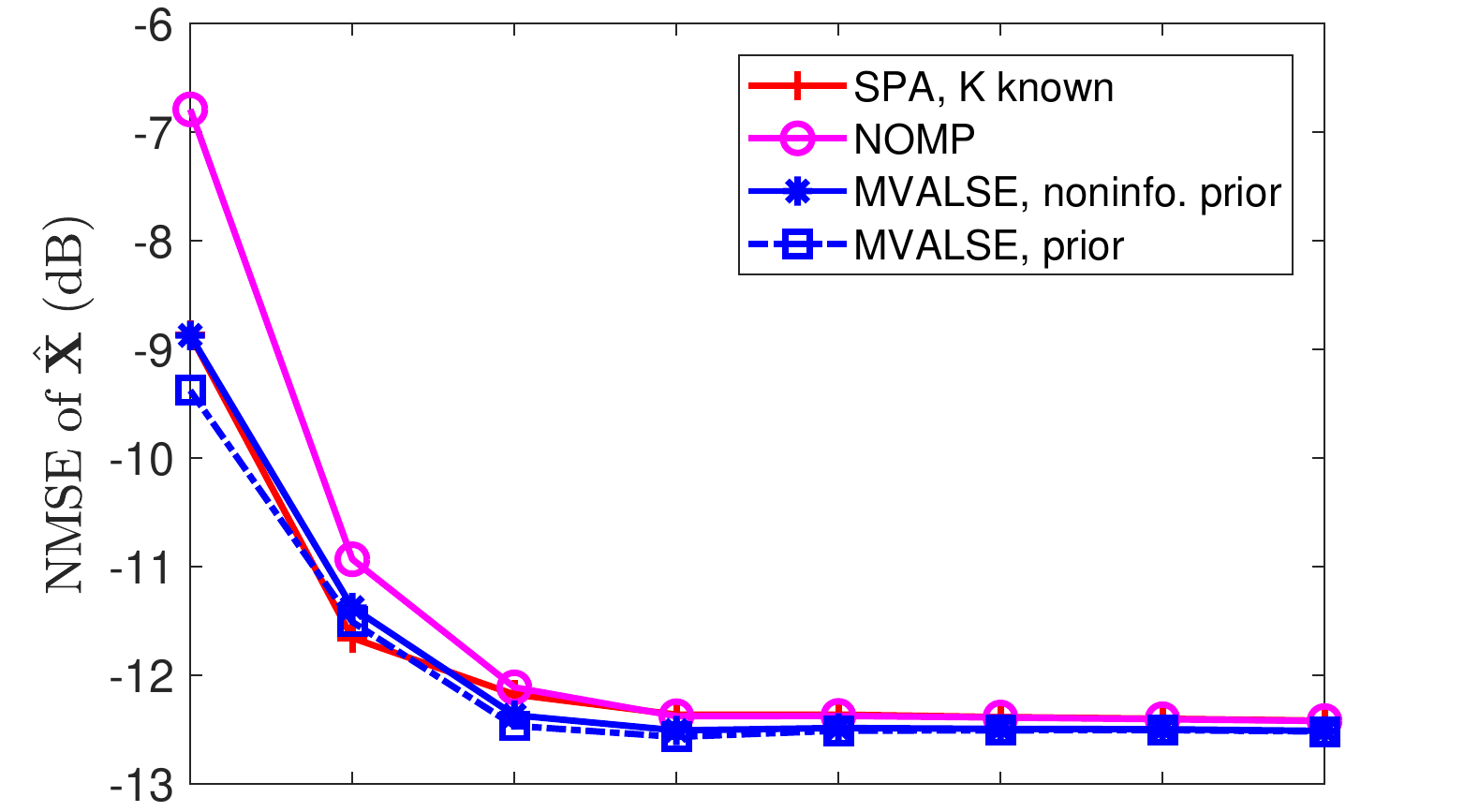}}
  \subfigure[]{
    \label{L-P} %% label for first subfigure
    \includegraphics[width=60mm]{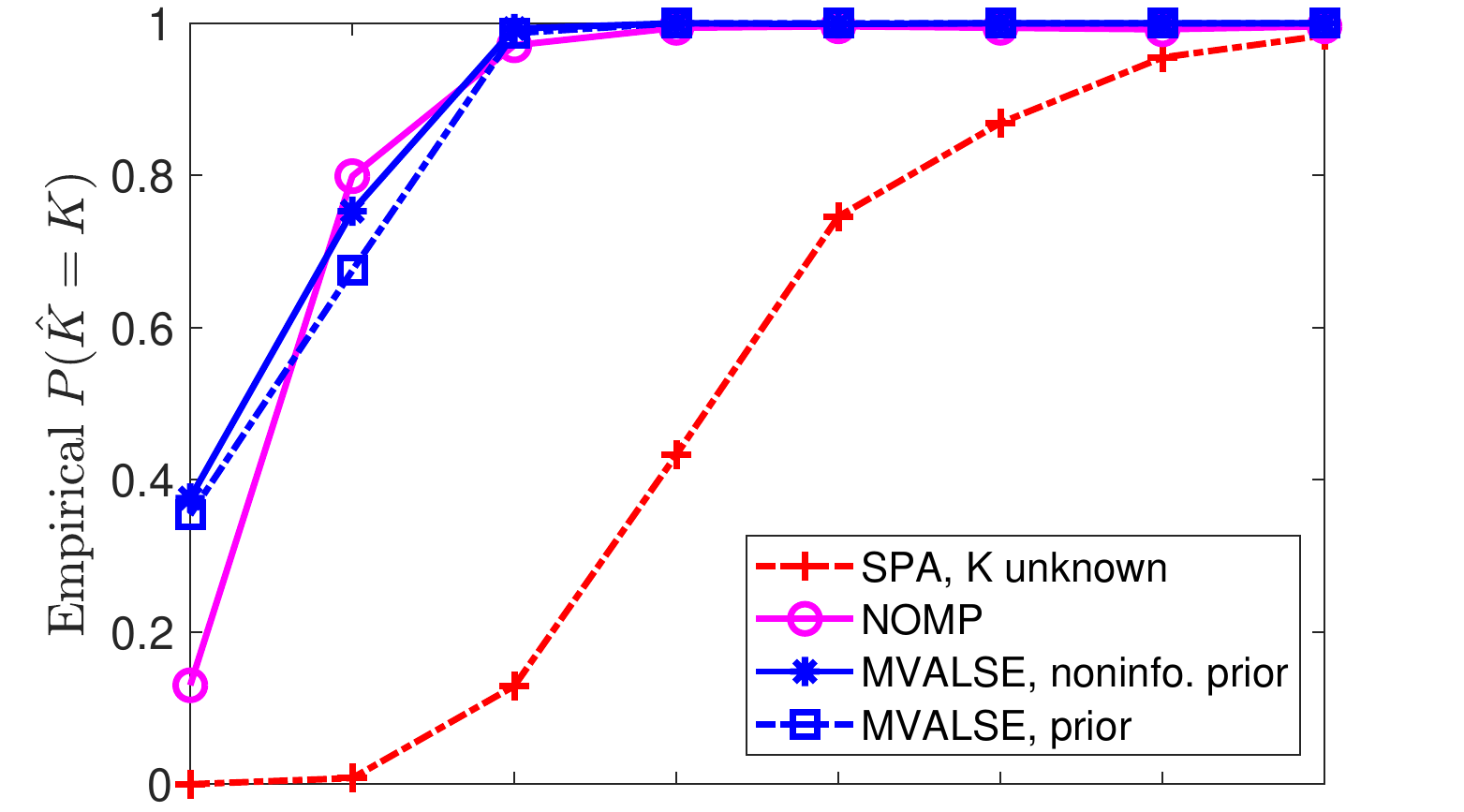}}
  \subfigure[]{
    \label{L-freq} %% label for second subfigure
    \includegraphics[width=60mm]{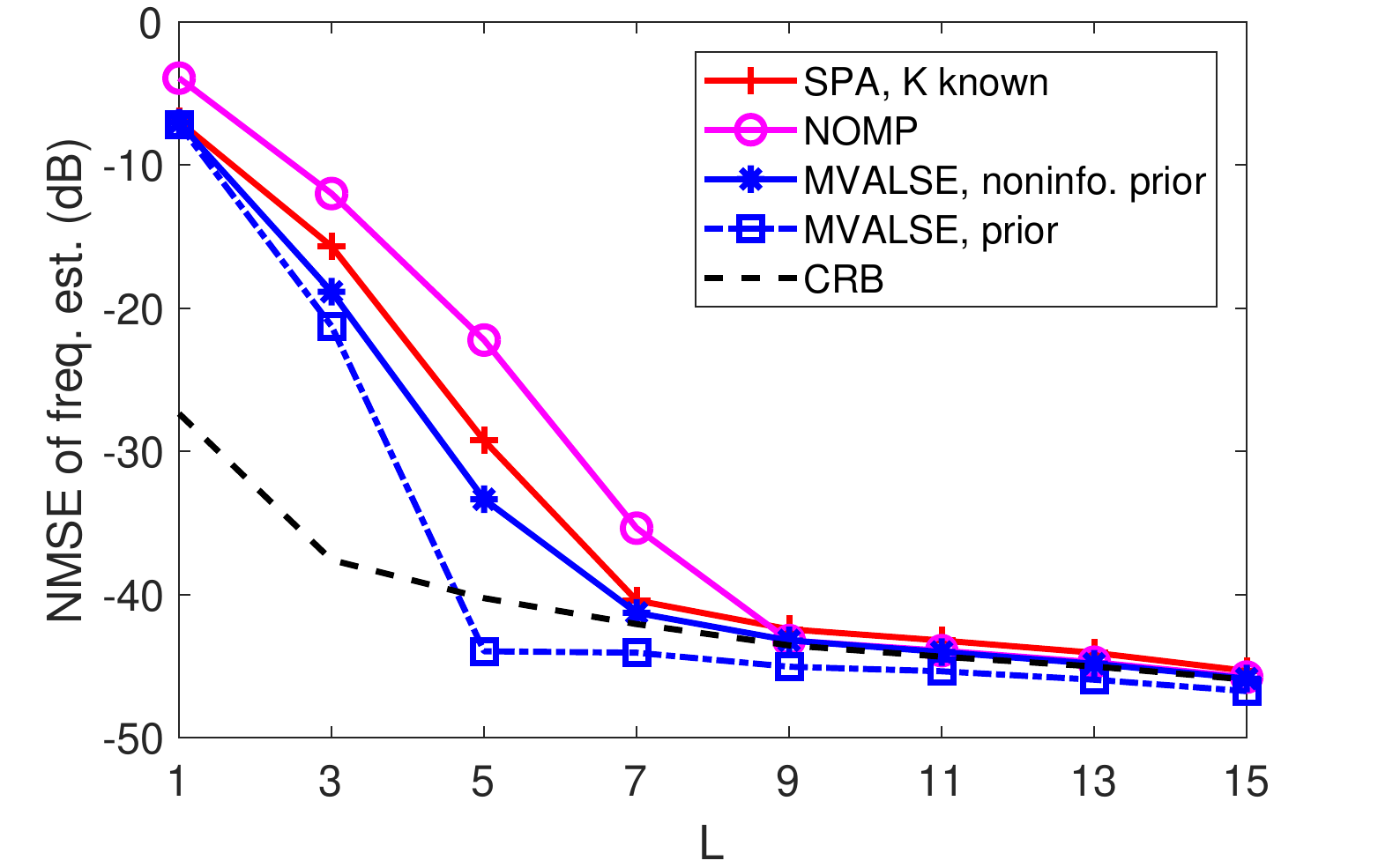}}
  \caption{Performance with varying snapshots $L$. Note that the SPA with $K$ unknown is not plotted in Fig. \ref{L-X} and Fig. \ref{L-freq} because of the poor performance.}
  \label{fig:subfig2} %% label for entire figure
\end{figure}

The estimation performance is examined by varying the number of snapshots $L$, shown in Fig. \ref{fig:subfig2}. As $L$ increase, the performances of all the algorithms improve. When $L\leq 3$, the signal reconstruction performance of the NOMP is worse than any other algorithms in Fig. \ref{L-X}. The reason is that the correct model order probability is not close to $1$, and the model order overestimate probability is only $1\%$, much smaller than the MVALSE methods shown in Table \ref{table}. For the frequency estimation error in Fig. \ref{L-freq}, all the algorithms except the MVALSE with prior approach to the CRLB as $L$ increases. The frequency estimation error of the MVALSE with prior is smaller than the CRLB when $L\geq 5$, which makes sense because prior information is utilized.

\section{Conclusion}
In this paper, the MVALSE algorithm is developed to solve the LSE with the MMVs, which is important for array signal processing. In addition, the von Mises distribution is encoded into the MVALSE algorithm, and the computation complexity of the MVALSE is analyzed. Numerical experiments show that increasing the number of snapshots improves the recovery performance, and the MVALSE algorithm is competitive compared to other state-of-the-art methods, especially when the prior distribution is imposed into the MVALSE algorithm.
\newpage{}
\bibliographystyle{IEEEbib}
\bibliography{strings,refs}

\end{document}